\documentclass[sensors,communication,accept,pdftex,moreauthors]{Definitions/mdpi} 

\firstpage{1} 
\pubvolume{24}
\issuenum{14}
\articlenumber{4621}
\pubyear{2024}
\copyrightyear{2024}
\externaleditor{Academic Editor: Carles Gomez}
\datereceived{6 June 2024} 
\daterevised{13 July 2024} 
\dateaccepted{16 July 2024} 
\datepublished{17 July 2024} 
\hreflink{https://\linebreak doi.org/10.3390/s24144621} 

\usepackage{threeparttable}

\Title{Opportunistic Sensor-Based Authentication Factors in and for the Internet of Things}

\TitleCitation{Opportunistic Sensor-Based Authentication Factors in and for the Internet of Things}

\Author{Marc Saideh $^{1,}$*\orcidA{}, Jean-Paul Jamont $^{2}$ and Laurent Vercouter $^{1}$}

\AuthorNames{Marc Saideh, Jean-Paul Jamont and Laurent Vercouter}

\AuthorCitation{Saideh, M.; Jamont, J.-P.; Vercouter, L.}

\address{%
$^{1}$ \quad INSA Rouen Normandie, Normandie Université, 
 LITIS UR 4108, 76000 Rouen, France; laurent.vercouter@insa-rouen.fr\\
$^{2}$ \quad Université Grenoble Alpes, LCIS, 26000 Valence, France; jean-paul.jamont@univ-grenoble-alpes.fr}

\corres{Correspondence: marc.saideh@insa-rouen.fr}

\abstract{Communication between connected objects in the Internet of Things (IoT) often requires secure and reliable authentication mechanisms to verify identities of entities and prevent unauthorized access to sensitive data and resources. Unlike other domains, IoT offers several advantages and opportunities, such as the ability to collect real-time data through numerous sensors. These data contains valuable information about the environment and other objects that, if used, can significantly enhance authentication processes. In this paper, we propose a~novel idea to building opportunistic sensor-based authentication factors by leveraging existing IoT sensors in a~system of systems approach. The objective is to highlight the promising prospects of opportunistic authentication factors in enhancing IoT security. We claim that sensors can be utilized to create additional authentication factors, thereby reinforcing existing object-to-object authentication mechanisms. By integrating these opportunistic sensor-based authentication factors into multi-factor authentication schemes, IoT security can be substantially improved. We demonstrate the feasibility and effectivenness of our idea through illustrative experiments in a~parking entry scenario, involving both mobile robots and cars, achieving high identification accuracy. We highlight the potential of this novel method to improve IoT security and suggest future research directions for formalizing and comparing our approach with existing techniques.}


\begin{document}

\section{Introduction}
The Internet of Things (IoT) is a~network of connected objects, including devices, vehicles, and~various other entities, often embedded with sensors and actuators. These objects collect and exchange data, enabling numerous applications that improve efficiency and automation. From~smart homes that improve comfort and optimize energy based on occupancy to industrial automation systems that optimize production processes, IoT is revolutionizing the way we interact with the world. However, the~increasing number of connected devices also brings significant security challenges, particularly in ensuring that communication between IoT objects is secure and~reliable.

One of the critical aspects of securing IoT communication is authentication, which is the process of verifying the identity of an~entity. Ensuring that communication occurs between authenticated entities prevents unauthorized access to sensitive data and resources. This is particularly important in~situations like object-to-object communication, where it is essential to confirm the identities of both objects before any exchange of information or interaction. The~diverse and distributed nature of IoT systems heightens the need for robust authentication mechanisms. These systems often operate under constraints that make traditional centralized authentication models less effective~\cite{el2019survey}.

\textls[5]{Over the last decade, extensive research has been conducted to address security challenges in IoT and to develop methods to protect against attacks~\cite{roman2013features, sicari2015security, meneghello2019iot, babun2021survey}. Authentication has been a~focal point of these efforts, as~robust authentication mechanisms are essential to mitigate the significant risks of unauthorized access to IoT networks~\cite{tawalbeh2020iot,meneghello2019iot}. Attackers could impersonate other objects, eavesdrop sensitive information and destroy trust relations between objects. Numerous approaches have been explored, including password-based mechanisms~\cite{shah2018authentication, renuka2019design, atwady2017survey} and advanced biometrics~\cite{yang2021biometrics}. Multi-Factor Authentication (MFA), which employs several factors to verify an~entity,  has emerged as a~more secure solution~\cite{ometov2018multi}. In~spite of these advancements, authentication in IoT remains a~significant challenge due to the dynamic and heterogeneous nature of IoT environments. IoT devices often have limited computational power and energy resources, making it difficult to implement complex authentication protocols. Additionally, the~diversity of devices and communication protocols in IoT networks adds another layer of complexity to the authentication~process.}

Despite these constraints and challenges, IoT also offers advantages and opportunities that can be leveraged to improve security measures. The~extensive deployment of sensors across various IoT environments provides a~rich source of data. These sensors continuously collect real-time information about their surroundings and the objects they interact with, which, if~utilized effectively, can significantly enhance existing security~protocols.

In this paper, we introduce a~novel concept aimed at reinforcing IoT authentication mechanisms by leveraging the existing sensors deployed in IoT environments within a~System of Systems (SoS) approach. We propose the idea of opportunistic sensor-based authentication factors, which utilizes the data already being collected by these sensors for purposes beyond their primary function. This innovative approach repurposes sensor data to create additional authentication factors, enhancing the security of IoT systems without the need for additional hardware~investments.

To demonstrate the potential of our idea, we conducted illustrative experiments in a~specific IoT scenario representing a~parking entry involving both mobile robots and cars. While the details of these experiments are elaborated in the subsequent sections, the~results underscore the feasibility and effectiveness of using opportunistic sensor-based authentication factors. This paper aims to highlight the potential of opportunistically leveraging existing IoT sensors to enhance security through innovative repurposing. By~integrating these opportunistic sensor-based factors into multi-factor authentication schemes, we can achieve a~higher level of security in IoT~environments.

The rest of this paper is organized as follows: Section~\ref{sec: background} provides a~background of relevant authentication methods in IoT that have been used so far, as~well as different types of authentication factors, while Section~\ref{sec: idea} explains the proposed idea and provides an~illustrative scenario. Next, in~Section~\ref{sec: experiments} we present the illustrative experiments conducted to demonstrate the proposed idea. At~the end, Sections~\ref{sec: discussion} and \ref{sec: conclusion} provide a~discussion and a~conclusion over the obtained results with some challenges and some hints for future~work.

\section{Background and Related~Work}
\label{sec: background}
Authentication is the process by which an~entity (human or machine) proves its identity by providing evidence that it is who it claims to be. These proofs are known as authentication factors. Authentication factors are categorized into three main groups~\cite{ometov2018multi}:
\begin{itemize}
\item Knowledge factor: something the user knows, such as a~password or PIN.
\item Possession factor: something the user has, such as a~security token.
\item Inherent factor: something the user is, such as a~fingerprint or facial recognition.
\end{itemize}

Given the diversity and sensitivity of IoT applications, reinforcing authentication mechanisms is critical. Several approaches have been proposed to enhance the robustness of these mechanisms. In~\cite{dasgupta2016toward}, the~authors proposed a~methodology to select the best authentication factors based on trust values assigned to the factors, as~well as the context, security requirements, and~associated risks. This approach emphasizes the dynamic and contextual nature of authentication in IoT~environments.

Context-aware authentication is another important method. The~authors of~\cite{miettinen2018revisiting} proposed an~authentication method that incorporates the context of the environment in which devices are authenticated. Context-aware authentication adapts the authentication process based on environmental factors such as location and time, thereby increasing security.
Location-based authentication utilizes the geographical location of the entity, such as GPS coordinates, to~verify its identity~\cite{zhang2012location}. This method adds an~additional layer of security by ensuring that authentication is tied to a~specific~location.

Building a~signature of entities involves using patterns related to the entities for authentication. These patterns can be based on behavioral or physical factors, which are part of the inherent factors. Authors in~\cite{liang2020behavioral} conducted a~survey on behavioral biometrics for continuous user authentication in IoT, highlighting how biometric data such as touchscreen dynamics, eye movements, and~other behaviors can strengthen authentication~mechanisms. 

Several studies have explored the use of sensor data to enhance security in IoT systems. For~instance, researchers have investigated the use of accelerometer data to authenticate smartphone users based on their phone usage characteristics~\cite{ouguz2022human}. Other studies have examined the potential of using environmental noise and light patterns as supplementary authentication factors~\cite{wang2020user}. While most studies have primarily focused on user authentication, some have investigated the use of Physically Unclonable Functions (PUFs), which leverage the unique physical properties of hardware components to generate cryptographic keys for device authentication~\cite{mall2022puf, rahim2018sensor}. These findings suggest the potential for extending such methods to other types of sensors. However, there is a~significant research gap in applying behavioral or physical signatures for object-to-object authentication in \mbox{IoT~environments.}

In summary, the extensive deployment of sensors in IoT environments presents an~opportunity to enhance authentication mechanisms by repurposing sensor data. The~following sections will introduce our proposed concept of opportunistic sensor-based authentication factors and show its potential through an~illustrative experimental case~study.

\section{Leveraging IoT Existing Systems for~Authentication}
\label{sec: idea}
In this section, we define the concept of opportunistic authentication factors, providing an~understanding of their characteristics and potential applications. To~clarify the practicality of these factors, we present an~illustrative scenario that effectively demonstrates their advantages. Following this, the~subsequent section presents illustrative experiments that reflect the proposed scenario, showing the potential of our new idea through \mbox{practical~examples.}

\subsection{Proposed Concept: Opportunistic Sensor-Based Authentication~Factors}
The increasing deployment of IoT devices has led to the widespread availability of various sensors embedded in these devices. These sensors continuously collect data for their primary functions, such as monitoring and control. The~core idea of opportunistic sensor-based authentication factors is to leverage the data collected by these existing IoT sensors, which are not typically used for authentication purposes, to~reinforce existing authentication mechanisms. This approach repurposes sensor data to serve as supplementary authentication factors, thereby enhancing the security of IoT systems without the need for additional hardware investments. By~leveraging existing IoT systems in a~SoS approach, we can integrate various independent systems to work together collectively, thereby enhancing the overall security infrastructure. This transformation of sensor data into authentication factors enables a~new level of authentication, making it more robust and~reliable.

IoT devices are equipped with a~variety of sensors, such as cameras, temperature sensors, accelerometers, and~light sensors. These sensors gather data continuously about their environment and the objects they interact with. For~example, temperature sensors can monitor heat levels, accelerometers can detect motion patterns, and~light sensors can measure the intensity of ambient~light.

To utilize sensor data for authentication, relevant features must be extracted from the raw data. This involves processing the data to identify unique characteristics that can be used to distinguish between different entities. For~example:
\begin{itemize}
    \item Visual data: shape patterns and~object recognition from camera streams.
    \item Temperature data: temperature fluctuations, heat distribution patterns, and~thermal signatures from temperature sensors.
    \item Motion data: movement patterns, acceleration profiles, and~behavioral signatures from accelerometers.
\end{itemize}

Once the relevant features are extracted, they are used to generate authentication factors. These factors can be integrated into multi-factor authentication schemes, providing an~additional layer of security. Moreover, using this kind of information can often be performed without the need for additional communication between objects or with the user. For~instance, in~a smart home environment, access to a~secure area could be authenticated not only by a~keypad entry code, but also by opportunistically recognizing the unique walking pattern of an~individual detected by floor pressure~sensors.

The opportunistic authentication factors are designed to complement existing authentication systems. They do not replace them, but rather enhance them by providing supplementary verification based on opportunistic data. This integration ensures that even if one authentication factor is compromised, the~system remains secure due to the presence of additional factors. The~proposed concept can be applied to various IoT scenarios. It is flexible enough to adapt to different types of sensors and data sources, making it suitable for a~wide range of applications. This flexibility introduces an~element of unpredictability, making it more challenging for attackers to predict the authentication factors that will be~used.

\subsection{Illustrative~Scenario}
\label{sec: scenario}
\begin{figure}[H]
    \includegraphics[width=0.99\textwidth]{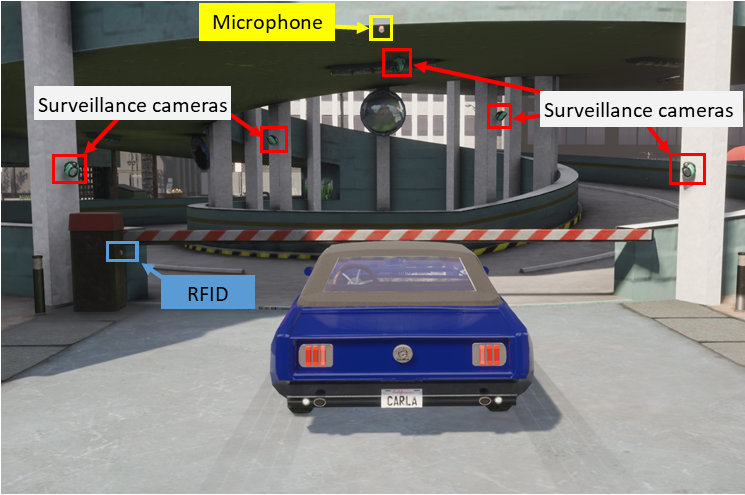}
    \caption{\label{fig:scenario}Illustrative scenario: parking~entry.}
\end{figure}

By illustrating the following scenario, we aim to highlight the feasibility and effectiveness of leveraging existing systems and sensor data in the authentication process. The~scenario involves a~private parking area with access control, where only authorized vehicles are allowed to enter. The~verification process of a~vehicle is performed at the entry to decide whether the vehicle should be allowed access. Three systems have been deployed in the parking area, each for specific functionalities. Figure~\ref{fig:scenario} (simulated with CARLA~\cite{dosovitskiy2017carla}) illustrates the scenario at the parking entrance. The~following provides a~description of the three existing~systems.

\unskip

\subsubsection{Description of the~Systems}
System 1: Barrier management system.
 Initially deployed to authenticate entering vehicles, the~barrier of the parking is managed using a~Radio Frequency Identification (RFID) scanner. Each authorized vehicle is equipped with an~RFID tag that is scanned at the entry to verify its identity.

System 2: Surveillance system. Deployed to ensure the overall safety of the parking area, including pedestrians and drivers, this system features strategically placed cameras that capture and record images and video footage of different areas. This enables real-time monitoring and enhances security measures.

System 3: Acoustic monitoring system. Deployed for safety reasons, this system detects abnormal sound patterns that may indicate incidents or aggressions. It includes microphones placed in specific locations, and~uses signal processing techniques to analyze recorded sounds and identify specific types of sounds or frequencies associated with alarms or other events to ensure security.

In this scenario, the~objective of our proposed idea is to reinforce the security of vehicle authentication. We aim to exploit the existing systems, initially deployed for other purposes than authentication, and~the data they collect through their embedded sensors. This exploitation provides new opportunistic authentication factors that enhance the currently deployed authentication~system.

\subsubsection{Utilizing Existing~Systems}
In this subsection, we explore how existing systems can be repurposed to enhance vehicle authentication. By~leveraging data from already deployed acoustic monitoring and surveillance systems, we aim to add new layers of~security.

\textbf{Acoustic monitoring system|Sensors: microphones.} While initially designed for safety applications, we propose using this system to strengthen vehicle authentication processes. Microphones placed at parking entrances can capture the engine sounds of vehicles waiting to enter. By~comparing these sounds with previously recorded engine sounds of authorized vehicles, we can help in verifying the identity of each vehicle. If~the captured sound is sufficiently similar to the known sound of an~authorized vehicle, the~sound factor is considered validated. This new recording can then be used for future reference. If~there is a~significant discrepancy between the sounds, it may indicate an~unauthorized vehicle or an~attempted security breach. 

\textbf{Surveillance system|Sensors: cameras.} Surveillance cameras, originally installed for security monitoring, can also enhance vehicle authentication. Various details can be extracted from images of vehicles at the parking entrance. These details can then be compared with information from authorized vehicles to verify identity. If~the captured details are sufficiently similar to those of an~authorized vehicle, the~visual factor is considered validated. Significant discrepancies may suggest an~unauthorized vehicle or a~potential security~threat.

\section{Illustrative~Experiments}
\label{sec: experiments}
Considering conditions similar to the described illustrative scenario, we evaluate the capacity of these existing IoT systems to reinforce the authentication process by creating signatures of the vehicles entering the parking, even by using simple processing methods. We take advantage of a~camera from the surveillance system and a~microphone from the acoustic monitoring system. We propose using the information perceived by these sensors to build signatures about the vehicles, that act as an~additional authentication~level.

\subsection{Measuring the Similarity of Engine~Sounds}
To compare two engine sounds, we were satisfied with employing a~simple yet effective signal processing method. Our objective is to demonstrate the feasibility of the proposed mechanism, focusing on the capability of identifying vehicles based on their engine sound rather than proposing the optimal signal processing method. We used the spectral centroid feature (Equation~(\ref{eq: SC})) to determine if two engine sounds are similar. It provides the average frequency or the center of mass of a~sound spectrum:
\begin{equation}
\label{eq: SC}
Spectral Centroid = \frac{\sum f(n)M(n)}{\sum M(n)}
\end{equation}
where $M(n)$ represents the magnitude or weighted frequency value of bin number $n$, and~$f(n)$ represents the center frequency at bin $n$.

With this feature, to~find the similarity or dissimilarity between two signals, we calculate the spectral centroid of each signal separately, then calculate the mean of each spectral centroid before computing the Euclidean distance between the two means (Equation~(\ref{eq: dist})):
\begin{equation}
distance = \sqrt{\sum(\overline{SC1} - \overline{SC2})^2}
\label{eq: dist}
\end{equation}
where $\overline{SC1}$ and $\overline{SC2}$ are the means of the first and second spectral centroids, respectively. As~the distance increases, the~dissimilarity between the two signals grows, while a~distance close to 0 indicates greater~similarity.

\subsection{Measuring the Similarity of Vehicle~Pictures}
In our experiments, a~series of vehicle images were captured against a~consistent background, and~the background was then removed to isolate the vehicles. To~illustrate the significance and potential of the features extracted from such information, we chose to use the color histogram of the captured images as a~feature for authentication. Although~color alone may not be a~uniquely distinguishing attribute, it can be crucial in identifying significant variations, particularly if a~potentially malicious vehicle has a~markedly different color. To~compare vehicles pictures, color histograms were generated from these isolated vehicle images, representing the distribution of colors for each vehicle. This comparison helps determine if two vehicle images represent the same vehicle based on the distribution of color data. We calculate the Bhattacharyya distance between two normalized color histograms (Equations~(\ref{eq: BC}) and (\ref{eq: BC2})):
\begin{equation}   
\begin{aligned}
    \label{eq: BC}
    D_B(P, Q) = -ln(BC(P,Q)) 
    \end{aligned}
\end{equation}
    \begin{equation}   
\begin{aligned}
    \label{eq: BC2}
    BC(P,Q) = \sum_x \sqrt{P(x)Q(x)}
\end{aligned}
\end{equation}
where $BC$ represents the Bhattacharyya coefficient, and~$P$ and $Q$ are the normalized color histograms of the two vehicle images. $BC$ measures the similarity between the two histograms. A~coefficient of 0 indicates that the histograms are identical, while higher values indicate greater~dissimilarity.

\subsection{Experimental~Setup}
The experiments were conducted on two different types of vehicles: cars and mobile robots, and~in two different environments: an~indoor environment for the mobile robots, where it is calm and there is not much noise when recording the engine sounds, and an~outdoor environment for the cars, where there is much more noise.  The~engines sounds were recorded from several distances from the vehicles to obtain realistic data and to detect if small distances can make a~significant difference in the comparison. Tests were repeated on several samples of different durations. For~example, one test was performed on samples of 5 s, another on samples of 2 s, and~another on 1 s samples. Regarding the second test, we captured multiple pictures of each vehicle, varying the camera angle slightly for each shot. The~collected data from cameras and microphones are used for feature extraction. The~extracted features are used to generate opportunistic authentication factors, integrated with the RFID data into the multi-factor authentication~mechanism.

\subsection{Results}
In this section, we present the results of the illustrative experiments conducted to evaluate the effectiveness of the proposed opportunistic sensor-based authentication factors in the described scenario. The~results are divided into two main categories: audio-based factors and visual-based~factors.

\subsubsection{Audio-Based Results|Engine Sounds}
\begin{figure}[H]
    \includegraphics[width=0.8\textwidth]{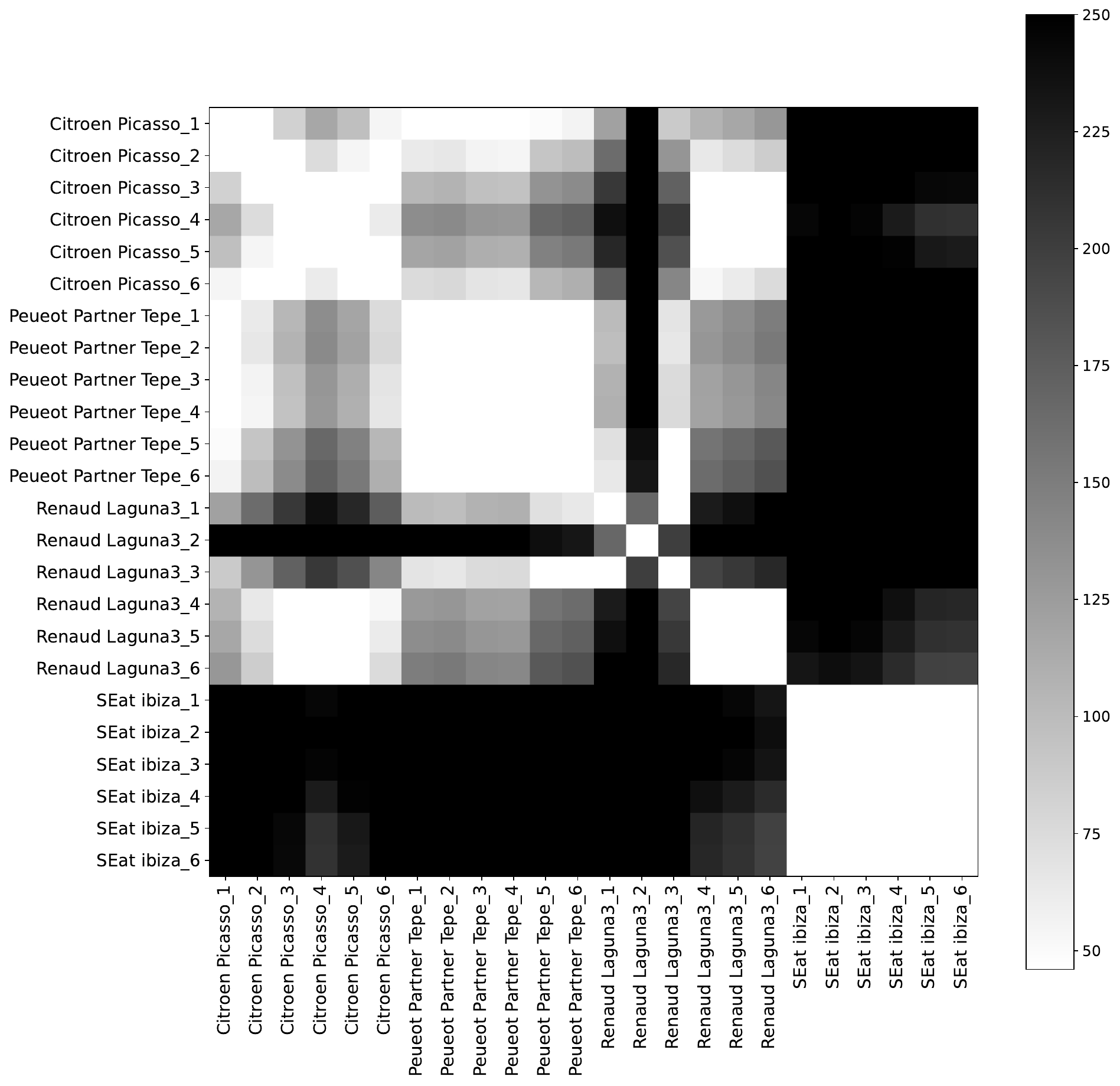}
    \caption{Distances between several samples of engine sounds of four different~cars.}
    \label{fig:distances_cars}
\end{figure}

\begin{figure}[H]
    \includegraphics[width=0.8\textwidth]{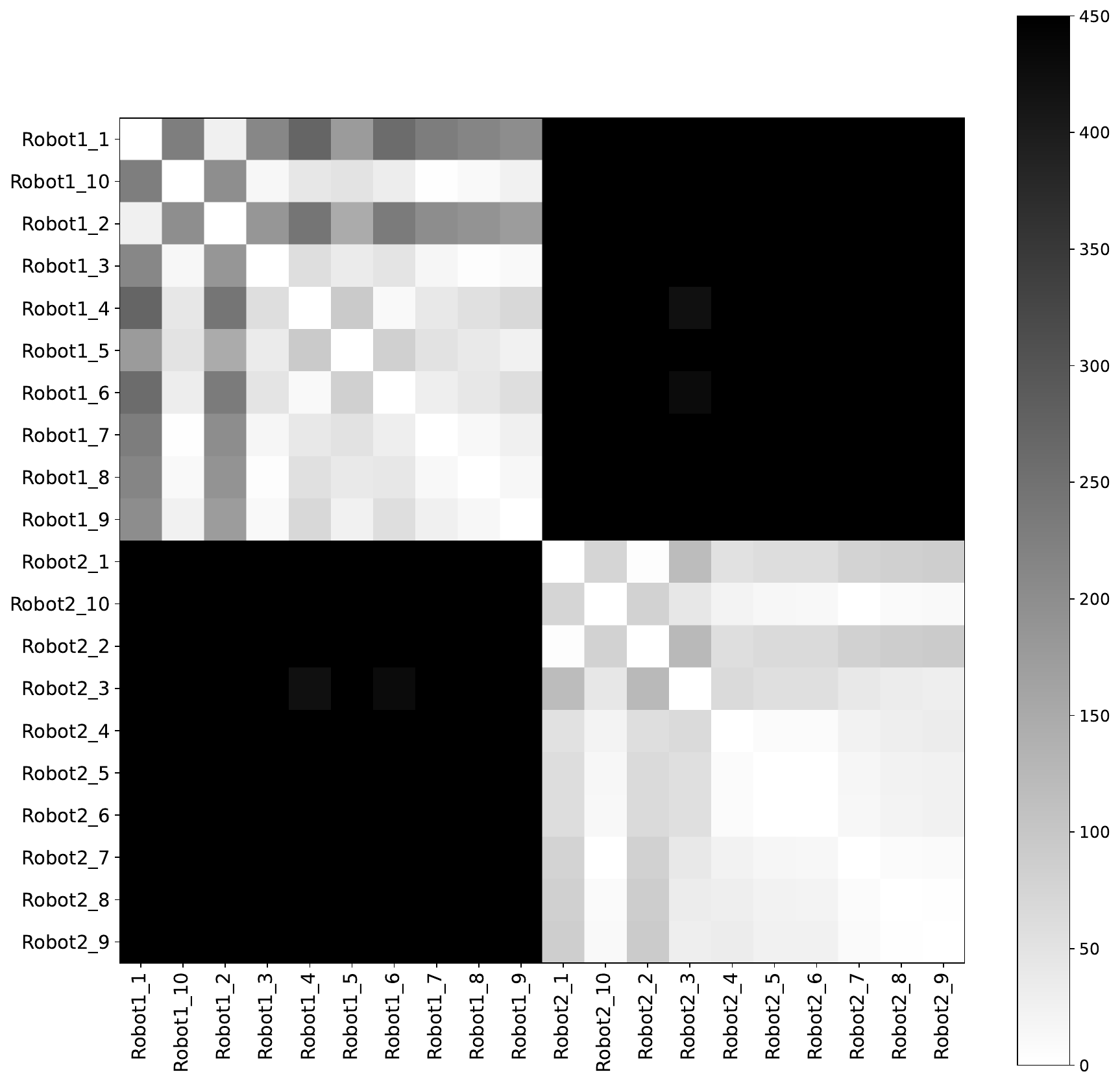}
    \caption{Distances between several samples of engine sounds of two different~robots.}
    \label{fig:distances_robots}
\end{figure}

The effectiveness of using engine sounds for vehicle authentication was measured using the spectral centroid feature and the Euclidean distance between spectral centroids of recorded engine sounds. The~results are summarized in Table~\ref{tab: avg_dist_sound} and Figures~\ref{fig:distances_cars} and \ref{fig:distances_robots}.

\begin{table}[H]
\caption{\label{tab: avg_dist_sound} Average distance between all the spectral centroids.}
\newcolumntype{C}{>{\centering\arraybackslash}X}
\begin{threeparttable}
\begin{tabularx}{\textwidth}{CCCCCCC}
\toprule
& \textbf{Robot 1} & \textbf{Robot 2} & \textbf{Car 1} & \textbf{Car 2} & \textbf{Car 3} & \textbf{Car 4} \\
\midrule
\textbf{Robot 1} & \textbf{43.6} & 467.8 & 951.3 & 1078.4 & 980.5 & 720.8 \\
\midrule
\textbf{Robot 2} & 467.8 & \textbf{38.5} & 514.3 & 635.5 & 422.8 & 290.5 \\
\midrule
\textbf{Car 1} & 951.3 & 514.3 & \textbf{17.2} & 131.1 & \textbf{\textcolor{red}{76.6}} & 223.9 \\
\midrule
\textbf{Car 2} & 1078.4 & 635.5 & 131.1 & \textbf{16.8} & 110.7 & 343.4 \\
\midrule
\textbf{Car 3} & 980.5 & 422.8 & \textbf{\textcolor{red}{76.6}} & 110.7 & \textbf{72.3} & 280.4 \\
\midrule
\textbf{Car 4} & 720.8 & 290.5 & 223.9 & 343.4 & 280.4 & \textbf{17.9} \\
\bottomrule
\end{tabularx}
	\noindent{\footnotesize{Note: \textbf{bold} values represent average distances that are true positives, while \textcolor{red}{red} ones indicate false positives.}}

\end{threeparttable}
\end{table}
\unskip

The conducted experiments resulted in a~100\% accuracy between the robots (indoor) and an~89\% accuracy between the cars (outdoor), utilizing a~fixed similarity score threshold of 100. Noting that the outdoor environment contains a~significant amount of noise, and Figures~\ref{fig:distances_cars} and \ref{fig:distances_robots} represent the symmetric distances between spectral centroids of different samples of engine sounds for cars and robots, respectively. In~these heatmaps, lighter cells represent more similar sounds, while darker cells indicate more different sounds. The~distinct clusters of lighter squares diagonally in each heatmap correspond to different vehicles, demonstrating that the engine sounds are consistent within the same vehicle and distinct between different vehicles. Table~\ref{tab: avg_dist_sound} represents a~symmetric distance matrix showing the average distance between all spectral centroids for all the tested vehicles (robots and cars). The~diagonal elements of the matrix represent the average distances between different spectral centroid samples of the same vehicle, while the off-diagonal elements represent the distances between different vehicles. From~the table, we observe one false positive where the average distance between samples from Car 1 and Car 3 is below 100, as~indicated in bold and red. However, it is evident that the average distance between spectral centroids of different vehicles is significantly greater than that between multiple spectral centroid samples from the same vehicle. All diagonal elements were under the fixed threshold, as~indicated in bold and black, demonstrating the potential for using such information for authentication~purposes.\vspace{-3pt}

\unskip

\subsubsection{Visual-Based Results|Color Histograms}
The effectiveness of using color histograms for vehicle authentication was measured by calculating the Bhattacharyya distance between the color histograms of captured vehicle images. The~results are summarized in Table~\ref{tab: avg_dist_color}.

The conducted experiments resulted in a~93.7\% accuracy when comparing color histograms of several samples for each car, utilizing a~fixed similarity score threshold of 0.2. For~the robots, a~100\% accuracy was achieved. Three of the four chosen cars for this experiment were selected to have similar color characteristics, minimizing potential bias in the results. Similar to the previous table, Table~\ref{tab: avg_dist_color} presents a~symmetric distance matrix illustrating the average distance between all color histograms for all the tested vehicles. The~average distance between color histograms of different vehicles is significantly greater than the average distance between color histograms of different picture samples from the same vehicle, often exceeding twice the value. The~only false negative result is observed in Car 3, where the average distance exceeds the fixed threshold, as~indicated in bold and red. This discrepancy can be attributed to slight variations in camera angles during the shots.

\begin{table}[H] 
\caption{\label{tab: avg_dist_color} Average distance between the color~histograms}
\newcolumntype{C}{>{\centering\arraybackslash}X}
\begin{threeparttable}
\begin{tabularx}{\textwidth}{CCCCCCC}
\toprule
& \textbf{Robot 1} & \textbf{Robot 2} & \textbf{Car 1} & \textbf{Car 2} & \textbf{Car 3} & \textbf{Car 4} \\
\midrule
\textbf{Robot 1} & \textbf{0.038} & 0.31 & - & - & - & - \\
\midrule
\textbf{Robot 2} & 0.31 & \textbf{0.042} & - & - & - & - \\
\midrule
\textbf{Car 1} & - & - & \textbf{0.069} & 0.222 & 0.382 & 0.376 \\
\midrule
\textbf{Car 2} & - & - & 0.222 & \textbf{0.145} & 0.342 & 0.348 \\
\midrule
\textbf{Car 3} & - & - & 0.382 & 0.342 & \textbf{\textcolor{red}{0.224}} & 0.382 \\
\midrule
\textbf{Car 4} & - & - & 0.376 & 0.348 & 0.382 & \textbf{0.165} \\
\bottomrule
\end{tabularx}
	\noindent{\footnotesize{Note: \textbf{bold} values represent average distances that are true positives, while \textcolor{red}{red} ones indicate false negatives.}}
\end{threeparttable}

\end{table}

The results from our illustrative experiments demonstrate the potential of opportunistically using sensors from neighboring systems to propose additional authentication factors as a~proof of concept (the code used for our illustrative experiments is available on GitHub: \url{https://github.com/Marc-SA/illustrative_exp_opportunistic_factors} (accessed on 16 July 2024)). The experiments using color histograms and spectral centroid for vehicle authentication are illustrative, showing that even simple methods can be used to build new authentication factors. Both methods highlight the potential of leveraging existing sensor data to enhance security. While color histograms can be processed efficiently and implemented using existing camera systems, and~the spectral centroid method can analyze engine sounds for authentication, both approaches may be sensitive to various environmental factors. Therefore, their use should be context-aware and ideally part of a~multi-factor authentication scheme to ensure robustness and~reliability.

In the following section, we discuss the implications of these results, the~potential benefits, and~the challenges associated with implementing opportunistic authentication factors in IoT~systems.

\section{Discussion}
\label{sec: discussion}
The concept of opportunistically leveraging existing IoT sensors to create new authentication factors is a~promising approach to enhancing the security of IoT systems. The~high accuracies achieved in our illustrative experiments with both audio-based and visual-based authentication demonstrate the feasibility and potential of this method. These results validate the integration of opportunistic factors with existing authentication mechanisms to create a~robust MFA system. There are several potential benefits to combining opportunistic~factors:

\begin{itemize}
    \item Increased security: By integrating existing authentication factors, as~RFID-based identification in the scenario, with~opportunistic factors, such as audio and visual data, the~overall security of the authentication system can be significantly enhanced. This multi-factor approach ensures that even if one authentication factor is compromised, additional factors provide robust verification, making unauthorized access much \mbox{more difficult.} 
    \item Reduced false positives: Utilizing multiple, diverse authentication factors helps reduce the likelihood of false positives. Each factor adds a~layer of verification, increasing the overall reliability of the authentication process.
    \item Enhanced robustness: A~multi-factor authentication system that leverages opportunistic data from existing sensors is more resilient against potential attacks. Attackers would need to compromise multiple authentication factors simultaneously, which is considerably more challenging than targeting a~single factor. Additionally, the~flexibility of opportunistic factors introduces an~element of unpredictability, complicating attackers' efforts as they would need to bypass multiple, distinct, and~not predetermined authentication factors.
\end{itemize}

\textls[5]{This method can be adapted to various types of IoT devices and environments by identifying useful data from different sensors and developing context-aware strategies for their use. By~quantifying the energy cost, security efficiency, and~latency of each sensor-based authentication factor, we can adapt the opportunistic method to other types of IoT sensors and applications. For~example, temperature sensors could detect environmental changes, motion detectors could identify unusual activities, and~in an~industrial IoT environment, vibration sensors could monitor equipment conditions to detect anomalies that suggest unauthorized access or tampering. Designers of IoT authentication protocols play a~crucial role in selecting appropriate sensors, integrating sensor data, and~implementing flexible authentication processes that enhance security across diverse applications. However, our experiments with sensor data from two environments showed context-dependence, with~indoor results outperforming outdoor ones. This outperformance is due to consistent lighting conditions, fewer weather-related disruptions, and~less background noise in indoor environments compared to outdoor settings. This highlights the importance of considering the context in which the sensors operate when evaluating the relevance of each~factor.}

Furthermore, the~ability to build a~large set of possible authentication factors provides the system with the flexibility to dynamically adapt to different contexts and threat scenarios. This adaptability allows the system to select the most relevant authentication factors for each specific situation, thus improving the overall security of the IoT environment. Additionally, thoughtful design and context-aware selection strategies are essential for effectively deploying this approach in diverse IoT environments, as~these strategies help balance resource demands and security~needs.

\section{Conclusions}
\label{sec: conclusion}
This study presents the novel concept of building opportunistic authentication factors in the IoT domain by leveraging data from existing sensors. Our goal was to demonstrate the potential of these opportunistic factors in enhancing the security and reliability of IoT authentication mechanisms without the need for additional hardware investments. Through illustrative experiments, we showed the feasibility of this approach by using simple methods to analyze data from existing sensors. These experiments validate the idea that opportunistic factors can serve as effective authentication elements, achieving high accuracy rates. The~results highlight the potential of integrating these opportunistic factors with existing ones to form a~robust MFA~system.

While there are numerous constraints and challenges inherent in IoT authentication mechanisms, our approach provides a~promising framework for addressing some of these issues. Indeed, by~creating a~set of possible authentication factors, this approach allows for dynamic selection based on current context and constraints. This selection of factors depends on several key criteria, including but not limited to resource availability, threat level (criticality of the interaction or the exchanged information), trust values, and~environmental conditions. For~example, when computing power or energy is limited, the~system can opt for less demanding factors, and~when resources are available, more complex factors can be employed. This adaptability enhances flexibility and interoperability in the IoT~environment.

To conclude, building new authentication factors in an~opportunistic way by leveraging existing IoT sensors is a~promising approach to enhancing IoT security. Future research should focus on extending this approach to a~wider range of sensor types and exploring the trade-offs between the cost, power consumption, and~security benefits of different authentication factors. Additionally, further studies are needed to validate the effectiveness of opportunistic authentication in diverse real-world scenarios and to develop methodologies for dynamically selecting the most relevant factors based on the context and threat~scenario.
\clearpage
%

\vspace{6pt} 
\authorcontributions{The work presented here was carried out in collaboration between all authors. Conceptualization, M.S., J.-P.J. and L.V.; methodology, M.S., J.-P.J. and L.V.; validation, M.S., J.-P.J. and L.V.; writing---original draft, M.S.; writing---review and editing, J.-P.J. and L.V. All authors have read and agreed to the published version of the~manuscript.}

\funding{This work is supported by the French National Research Agency (ANR) in the framework of the project MaestrIoT ANR-21-CE23-0016.}
 
 \institutionalreview{Not applicable}

\informedconsent{Not applicable}

\dataavailability{The data presented in this study are available upon request from the corresponding author.} 

\conflictsofinterest{The authors declare no conflicts of interest.} 


\abbreviations{Abbreviations}{
The following abbreviations are used in this manuscript:\\

\noindent 
\begin{tabular}{@{}ll}
IoT & Internet of Things\\
RFID & Radio Frequency Identification\\
SoS & System of Systems\\
PUF & Physical Unclonable Function\\
MFA & Multi-Factor Authentication 
\end{tabular}
}


\begin{adjustwidth}{-\extralength}{0cm}

\reftitle{References}

\PublishersNote{}
\end{adjustwidth}
\end{document}